\documentstyle[12pt]{article}
\input psfig
\catcode`\@=11
\long\def\@makefntext#1{
\protect\noindent \hbox to 3.2pt {\hskip-.9pt  
$^{{\ninerm\@thefnmark}}$\hfil}#1\hfill}                

\def\@makefnmark{\hbox to 0pt{$^{\@thefnmark}$\hss}}  
        
\def\ps@myheadings{\let\@mkboth\@gobbletwo
\def\@oddhead{\hbox{}
\rightmark\hfil\ninerm\thepage}   
\def\@oddfoot{}\def\@evenhead{\ninerm\thepage\hfil
\leftmark\hbox{}}\def\@evenfoot{}
\def\sectionmark##1{}\def\subsectionmark##1{}}

\renewcommand{\thefootnote}{\fnsymbol{footnote}}

\newcounter{sectionc}\newcounter{subsectionc}\newcounter{subsubsectionc}
\renewcommand{\section}[1] {\vspace*{0.6cm}\addtocounter{sectionc}{1} 
\setcounter{subsectionc}{0}\setcounter{subsubsectionc}{0}\noindent 
        {\normalsize\bf\thesectionc. #1}\par\vspace*{0.4cm}}
\renewcommand{\subsection}[1] {\vspace*{0.6cm}\addtocounter{subsectionc}{1} 
        \setcounter{subsubsectionc}{0}\noindent 
        {\normalsize\it\thesectionc.\thesubsectionc. #1}\par\vspace*{0.4cm}}
\renewcommand{\subsubsection}[1]
{\vspace*{0.6cm}\addtocounter{subsubsectionc}{1}
        \noindent {\normalsize\rm\thesectionc.\thesubsectionc.\thesubsubsectionc. 
        #1}\par\vspace*{0.4cm}}

\newcounter{appendixc}
\newcounter{subappendixc}[appendixc]
\newcounter{subsubappendixc}[subappendixc]

\renewcommand{\appendix}[1] {\vspace*{0.6cm}
        \refstepcounter{appendixc}
        \setcounter{figure}{0}
        \setcounter{table}{0}
        \setcounter{equation}{0}
        \renewcommand{\thefigure}{\Alph{appendixc}.\arabic{figure}}
        \renewcommand{\thetable}{\Alph{appendixc}.\arabic{table}}
        \renewcommand{\theappendixc}{\Alph{appendixc}}
        \renewcommand{\theequation}{\Alph{appendixc}.\arabic{equation}}
        \noindent{\bf Appendix \theappendixc #1}\par\vspace*{0.4cm}}

\def\abstracts#1{{
        \centering{\begin{minipage}{12.2truecm}\footnotesize\baselineskip=12pt\noindent
        \centerline{\footnotesize ABSTRACT}\vspace*{0.3cm}
        \parindent=0pt #1
        \end{minipage}}\par}} 

\renewenvironment{thebibliography}[1]
        {\begin{list}{\arabic{enumi}.}
        {\usecounter{enumi}\setlength{\parsep}{0pt}
\setlength{\leftmargin 1.25cm}{\rightmargin 0pt}
         \setlength{\itemsep}{0pt} \settowidth
        {\labelwidth}{#1.}\sloppy}}{\end{list}}

\newcounter{itemlistc}
\newcounter{romanlistc}
\newcounter{alphlistc}
\newcounter{arabiclistc}

\newcommand{\fcaption}[1]{
        \refstepcounter{figure}
        \setbox\@tempboxa = \hbox{\footnotesize Fig.~\thefigure. #1}
        \ifdim \wd\@tempboxa > 6in
           {\begin{center}
        \parbox{6in}{\footnotesize\baselineskip=12pt Fig.~\thefigure. #1}
            \end{center}}
        \else
             {\begin{center}
             {\footnotesize Fig.~\thefigure. #1}
              \end{center}}
        \fi}

\newcommand{\tcaption}[1]{
        \refstepcounter{table}
        \setbox\@tempboxa = \hbox{\footnotesize Table~\thetable. #1}
        \ifdim \wd\@tempboxa > 6in
           {\begin{center}
        \parbox{6in}{\footnotesize\baselineskip=12pt Table~\thetable. #1}
            \end{center}}
        \else
             {\begin{center}
             {\footnotesize Table~\thetable. #1}
              \end{center}}
        \fi}

\def\@citex[#1]#2{\if@filesw\immediate\write\@auxout
        {\string\citation{#2}}\fi
\def\@citea{}\@cite{\@for\@citeb:=#2\do
        {\@citea\def\@citea{,}\@ifundefined
        {b@\@citeb}{{\bf ?}\@warning
        {Citation `\@citeb' on page \thepage \space undefined}}
        {\csname b@\@citeb\endcsname}}}{#1}}

\newif\if@cghi
\def\cite{\@cghitrue\@ifnextchar [{\@tempswatrue
        \@citex}{\@tempswafalse\@citex[]}}
\def\citelow{\@cghifalse\@ifnextchar [{\@tempswatrue
        \@citex}{\@tempswafalse\@citex[]}}
\def\@cite#1#2{{$\null^{#1}$\if@tempswa\typeout
        {IJCGA warning: optional citation argument 
        ignored: `#2'} \fi}}

 1
 1
 1

\font\ninerm=cmr9

\catcode`\@=11
\def\marginnote#1{}
\newcount\hour
\newcount\minute
\newtoks\amorpm
\hour=\time\divide\hour by60
\minute=\time{\multiply\hour by60 \global\advance\minute by-
\hour}
\edef\standardtime{{\ifnum\hour<12 \global\amorpm={am}%
    \else\global\amorpm={pm}\advance\hour by-12 \fi
    \ifnum\hour=0 \hour=12 \fi
    \number\hour:\ifnum\minute<100\fi\number\minute\the\amorpm}}
\edef\militarytime{\number\hour:\ifnum\minute<100\fi\number\minute}
\def\draftlabel#1{{\@bsphack\if@filesw {\let\thepage\relax
  \xdef\@gtempa{\write\@auxout{\string
    \newlabel{#1}{{\@currentlabel}{\thepage}}}}}\@gtempa
    \if@nobreak \ifvmode\nobreak\fi\fi\fi\@esphack}
     \gdef\@eqnlabel{#1}}
\def\@eqnlabel{}
\def\@vacuum{}
\def\draftmarginnote#1{\marginpar{\raggedright\scriptsize\tt#1}}
\def\draft{\oddsidemargin -.5truein
        \def\@oddfoot{\sl preliminary draft \hfil
        \rm\thepage\hfil\sl\today\quad\militarytime}
        \let\@evenfoot\@oddfoot \overfullrule 3pt
        \let\label=\draftlabel
        \let\marginnote=\draftmarginnote
   
\def\@eqnnum{(\theequation)\rlap{\kern\marginparsep\tt\@eqnlabel}%
\global\let\@eqnlabel\@vacuum}  }
\def\preprint{\twocolumn\sloppy\flushbottom\parindent 1em
        \leftmargini 2em\leftmarginv .5em\leftmarginvi .5em
        \oddsidemargin -.5in    \evensidemargin -.5in
        \columnsep 15mm \footheight 0pt
        \textwidth 250mmin      \topmargin  -.4in
        \headheight 12pt \topskip .4in
        \textheight 175mm
        \footskip 0pt
        
\def\@oddhead{\thepage\hfil\addtocounter{page}{1}\thepage}
        \let\@evenhead\@oddhead \def\@oddfoot{} \def\@evenfoot{} 
}
\def\titlepage{\@restonecolfalse\if@twocolumn\@restonecoltrue\onecolumn
     \else \newpage \fi \thispagestyle{empty}\c@page\z@
        \def\thefootnote{\fnsymbol{footnote}} }
\def\endtitlepage{\if@restonecol\twocolumn \else  \fi
        \def\thefootnote{\arabic{footnote}}
        \setcounter{footnote}{0}}  
\catcode`@=12
\relax
\def\be{\begin{equation}}
\def\ee{\end{equation}}
\def\bea{\begin{eqnarray}}
\def\eea{\end{eqnarray}}
\def\simlt{\stackrel{<}{{}_\sim}}
\def\simgt{\stackrel{>}{{}_\sim}}

\def\NPB#1#2#3{{\it Nucl.~Phys.} {\bf{B#1}} (19#2) #3}
\def\PLB#1#2#3{{\it Phys.~Lett.} {\bf{B#1}} (19#2) #3}
\def\PRD#1#2#3{{\it Phys.~Rev.} {\bf{D#1}} (19#2) #3}
\def\PRL#1#2#3{{\it Phys.~Rev.~Lett.} {\bf{#1}} (19#2) #3}

\def\HPA#1#2#3{{\it Helv.~Phys.~Acta} {\bf#1} (19#2) #3}
\def\AP#1#2#3{{\it Ann.~Phys.} {\bf#1} (19#2) #3}
\def\MPLA#1#2#3{{\it Mod.~Phys.~Lett} {\bf A#1} (19#2) #3}

\def\mst11{m_{\;\widetilde{t}_{1}}}

\def\mst22{m_{\;\widetilde{t}_{2}}}
\def\mst12{m_{\;\widetilde{t}_{1,2}}}

\def\msb11{m_{\;\widetilde{b}_{1}}}
\def\mstop{m_{\;\widetilde{t}}}
\def\msb22{m_{\;\widetilde{b}_{2}}}
\def\msb12{m_{\;\widetilde{b}_{1,2}}}

\def\mtilde2{\widetilde{m}^{2}}

\relax
\begin{document}

\begin{flushright}
IEM-FT-152/96 \\
hep--ph/9703326\\
\end{flushright}

\vspace{3cm}

\centerline{\normalsize\bf ELECTROWEAK PHASE TRANSITION AND}
\baselineskip=22pt
\centerline{\normalsize\bf BARYOGENESIS IN THE MSSM~\footnote{
To appear in the  Proceedings of the Workshop on {\it The Higgs
puzzle-- What can we learn from LEP II, LHC, NLC and FMC?}, Ringberg
Castle, Germany, December 8-13, 1996. Ed. B.~Kniehl, World
Scientific, Singapore.}}

\vfill
\vspace*{0.6cm}
\centerline{\footnotesize MARIANO QUIROS}
\baselineskip=13pt
\centerline{\footnotesize\it Instituto de Estructura de la
Materia (CSIC), Serrano 123}
\baselineskip=12pt
\centerline{\footnotesize\it 28006-Madrid, Spain}
\centerline{\footnotesize E-mail: quiros@pinar1.csic.es}
\vspace*{0.3cm}

\vspace*{0.9cm}
\abstracts{We have analyzed baryogenesis in the MSSM for the
light stop scenario, where the phase transition is strong
enough first order. We have found that enough baryon asymmetry
can be generated provided that the phase of $\mu$ be $\simgt$
0.01. Constraints from the electric dipole moment of the neutron
enforce the first and second generation squarks to have masses
${\cal O}$(few) TeV.}
 
\normalsize\baselineskip=15pt
\setcounter{footnote}{0}
\renewcommand{\thefootnote}{\alph{footnote}}

\section{Introduction}

The option of generating the cosmological baryon
asymmetry~\cite{baryogenesis} at the electroweak phase
transition is not necessarily the one chosen by Nature, but it
is certainly fascinating, and has recently deserved a lot of
attention~\cite{reviews}. At the quantitative level, the
Standard Model (SM) meets the basic requirements for a successful
implementation of this scenario due to the presence of anomalous
processes~\cite{anomaly}.
However, the electroweak phase transition is too weakly first order
to assure the preservation of the generated baryon asymmetry at
the electroweak phase transition~\cite{first},
as perturbative~\cite{improvement,twoloop} and
non-perturbative~\cite{nonpert} analyses have shown. On the other
hand, CP-violating processes are suppressed by powers of $m_f/M_W$,
where $m_f$ are the light-quark masses.
These suppression factors are sufficiently strong
to severely restrict the possible baryon number generation~\cite{fs,huet}.
Therefore, if the baryon asymmetry is generated at the electroweak phase
transition, it will require the presence of new physics at the 
electroweak scale.

Low energy supersymmetry is a well motivated possibility, and it is
hence highly interesting to test under which conditions there is
room for electroweak baryogenesis in this
scenario~\cite{early,mariano1,mariano2}.
It was recently shown~\cite{CQW}
that the phase transition can be sufficiently strongly first
order in a restricted region of parameter space: 
The lightest stop
must be lighter than the top quark, the ratio of vacuum expectation
values $\tan\beta \simlt 3$, while the lightest Higgs must be at the
reach of LEP2.   
Similar results were independently obtained by
the authors of Ref.~\cite{Delepine}.
These results have been confirmed by explicit sphaleron calculations
in the Minimal Supersymmetric Standard Model
(MSSM)~\cite{MOQ}, while two-loop calculations have the general
tendency to strengthen the phase transition~\cite{CEQW,JoseR} thus making the
previous bounds very conservative ones.
On the other hand, the MSSM
contains, on top of the Cabbibo-Kobayashi-Maskawa 
matrix phase, additional sources of CP-violation
and can account for the observed baryon asymmetry.
New CP-violating phases can arise
from the soft supersymmetry breaking parameters associated with
the  stop mixing angle.  

In this talk I will review the computation of the baryon asymmetry 
and the strength of the first order phase transition in the MSSM. 
I will identify the region
in the supersymmetric parameter space where baryon asymmetry is
consistent with the observed value and,
furthermore, it is not washed out inside the bubbles after the
phase transition. 

\section{The phase transition in the MSSM}

A strongly first order electroweak phase transition
can  be achieved in the presence of a top squark
lighter than the top quark~\cite{CQW}. In order to naturally
suppress its contribution to the parameter $\Delta\rho$, and hence
preserve a good agreement with the precision measurements at LEP,
it should be mainly right handed. This can be achieved if the left
handed stop soft supersymmetry breaking mass $m_Q$
is much larger than $M_Z$.
For moderate mixing,
the lightest stop mass is then approximately given by
\be
\label{masastop}
\mstop^2 = m_U^2 + D_R^2 + m_t^2(\phi) \left( 1  -
\frac{\widetilde{A}_t^2}{m_Q^2}
\right)
\ee
where $\widetilde{A}_t = A_t - \mu/\tan\beta$ is the
particular combination appearing in the off-diagonal terms of
the left-right stop squared mass matrix and $m_U^2$ is
the soft supersymmetry breaking squared mass parameter
of the right handed stop.  

In order to overcome the Standard Model
constraints, the stop contribution must be large.
The stop contribution strongly depends
on the value of $m_U^2$, which must be small in magnitude, and
negative, in order to induce a sufficiently strong first order phase
transition. Indeed, large stop contributions
are always associated with small values of the right handed stop
plasma mass
\begin{equation}
m^{\rm eff}_{\;\widetilde{t}} = -\widetilde{m}_U^2 + \Pi_R(T)
\label{plasm}
\end{equation}
where $\widetilde{m}_U^2 = - m_U^2$, $\Pi_R(T) \simeq 4 g_3^2
T^2/9+h_t^2/6[2-
\widetilde{A}_t^2/m_Q^2]T^2$~\cite{CQW,CE} is the finite
temperature
self-energy contribution to the right-handed
squarks, and $h_t$ and $g_3$  are the
top quark Yukawa and strong gauge couplings, respectively.
We are considering heavy (decoupled from the
thermal bath) gluinos. For light gluinos, their contribution
to the squark self-energies,
$2g_3^2 T^2/9$, should be added to 
$\Pi_R(T)$~\cite{mariano1}.
Moreover, the trilinear mass term, $\widetilde{A}_t$,
must be $\widetilde{A}_t^2 \ll m_Q^2$
in order to avoid the suppression of  the stop contribution
to $v(T_c)/T_c$. The dependence of the order parameter
$v(T_c)/T_c$ on $\widetilde{m}_U$ is illustrated in Fig.~1a
where we plot it as a function of the light stop mass
(\ref{masastop}). We see from it a dramatic increase in
$v(T_c)/T_c$ as $\widetilde{m}_U$ increases.

\begin{figure}[htb]
\centerline{
\psfig{figure=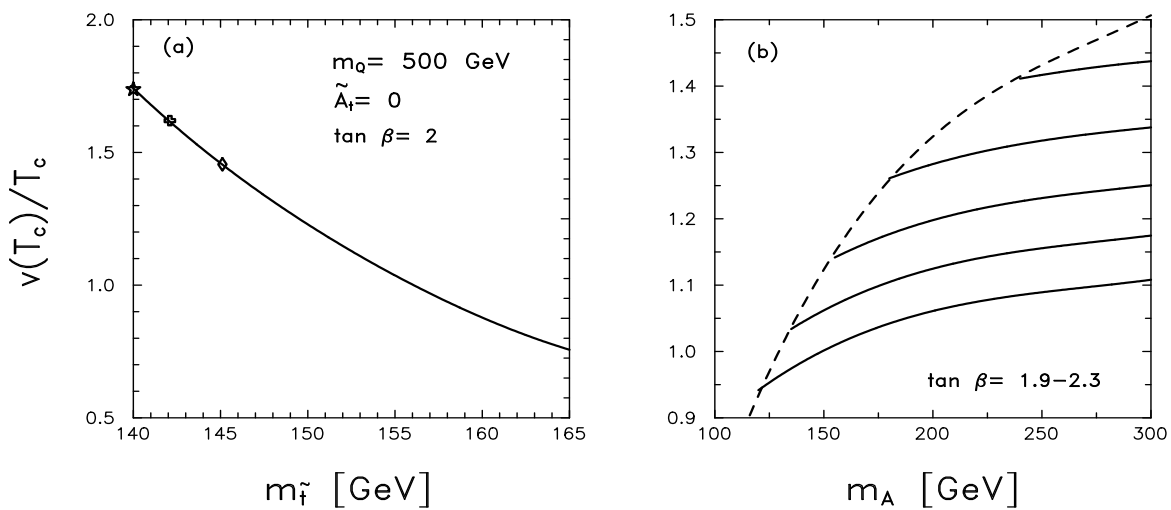,height=7.5cm,width=6cm,bbllx=9.cm,bblly=3.5cm,bburx=18.5cm,bbury=16.5cm}}
\caption{{\bf a)} Plot of $v(T_c)/T_c$ as a function of $m_{\widetilde{\; t}}$
for $m_Q=m_A=500$ GeV, $\widetilde{A}_t=0$ and $\tan\beta=2$. The
diamond denotes $\widetilde{m}_U=\widetilde{m}_U^{\rm  crit}$, Eq.~(3). 
{\bf b)} Plots of $v(T_c)/T_c$ as functions of $m_A$ (solid lines)  for
$\tan\beta=1.9\ ({\rm upper\ line})-2.3\ ({\rm lower line})$, 
step=0.1, $m_Q=500$ GeV and $\widetilde{m}_U=
\widetilde{m}_U^{\rm  crit}$. The dashed line  corresponds to the
experimental bound $m_h=m_h^{\rm  exp}$.}
\end{figure}

Although large values of $\widetilde{m}_U$, of order of the critical
temperature,
are useful to achieve a strongly first order phase transition, they may
also induce charge and color breaking minima. Indeed, if the
effective plasma mass at the critical temperature vanished, the
universe would be driven to a charge and color breaking minimum at
$T \geq T_c$~\cite{CQW}. A conservative
bound on $\widetilde{m}_U$ may be obtained by demanding that
the electroweak symmetry breaking minimum be lower than any
color-breaking minima induced by the presence of $\widetilde{m}_U$ at
zero temperature, which yields the condition
\begin{equation}
\widetilde{m}_U\simlt \widetilde{m}_U^{\rm  crit}
\equiv \left(\frac{m_h^2 v^2 g_3^2}{12}\right)^{1/4}.
\label{colorbound}
\end{equation}
It can be shown that this condition is sufficient to prevent dangerous
color breaking minima at zero and finite temperature for any value of
the mixing parameter $\widetilde{A}_t$ \cite{CQW}.
In this work, we shall use
this conservative bound.

Fig.~1a corresponds to a large value of the mass of the
pseudoscalar Higgs, for which the strength of
the phase transition is maximized~\cite{mariano2}.
However, for the purpose
of generating the baryon asymmetry, as we will see in the next
section, smaller values of $m_A$ should be used. In Fig.~1b we
present plots of $v(T_c)/T_c$ as a function of $m_A$ for
different values of $\tan\beta$. Every line stops at a lower
value of $m_A$, where the experimental LEP bound on the Higgs
mass is met. The region to the left of the dashed line in Fig.~1b is
excluded by LEP searches of the Higgs boson.

\begin{figure}[htb]
\centerline{
\psfig{figure=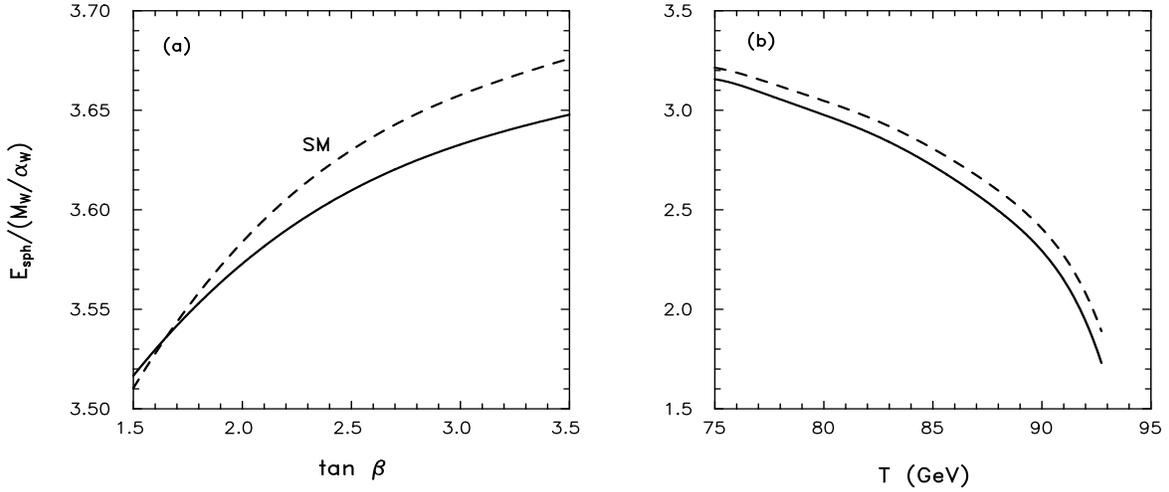,height=7.5cm,width=6cm,bbllx=9.cm,bblly=3.5cm,bburx=18.5cm,bbury=16.5cm}}
\caption{For $m_Q=m_A=500$ GeV, $\widetilde{A}_t=0$ and
$\widetilde{m}_U=\widetilde{m}_U^{\rm  crit}$: {\bf a)} $E_{\rm
sph}^{\rm MSSM}(0)$ (solid line) as a function of $\tan\beta$. The
dashed line is $E_{\rm sph}^{\rm SM}(0)$ for a Higgs mass equal to
$m_{\rm eff}$, Eq.~(6). {\bf b)} $E_{\rm sph}^{\rm MSSM}(T)$ for
$\tan\beta=2$ (solid line). The dashed line denotes a plot of $E_{\rm
sph}^{\rm MSSM}(0)v(T)/v$.}
\end{figure}

The requirement of not washing out, after the phase transition
the previously generated baryon asymmetry provides the 
condition~\cite{BKS}$ E_{\rm sph}(T_c)/T_c \simgt 45$, 
which translates, in the Standard Model, into the condition
\be
\label{condition}
\frac{v(T_c)}{T_c} \simgt 1.
\ee
In the MSSM the condition (\ref{condition}) should hold provided
$E_{\rm sph}^{\rm MSSM}(T_c) \sim E_{\rm sph}^{\rm
SM}(T_c)$. In particular this will hold if the scaling law
\be
\label{scaling} 4 
E_{\rm sph}^{\rm MSSM}(T_c)=E_{\rm sph}^{\rm MSSM}(0)
\frac{v(T_c)}{v}
\ee
is approximately satisfied, and at zero temperature
$
E_{\rm sph}^{\rm MSSM}\sim E_{\rm sph}^{\rm SM}(m_{\rm
eff}) 
$, 
where 
\be
m_{\rm eff}^2=\sin^2(\alpha-\beta)m_h^2+\cos^2(\alpha-\beta)m_H^2 
\ee 
$m_{h,H}$ being the light/heavy CP-even mass eigenstates,
and $\alpha$ the mixing angle in the Higgs sector, where all
radiative corrections effects corresponding to the chosen
supersymmetric parameters have been incorporated. 

In Fig.~2a we
compare $E_{\rm sph}^{\rm MSSM}$ (solid line) 
with $E_{\rm sph}^{\rm SM}$ (dashed line)
for a Higgs mass equal to $m_{\rm eff}$. In Fig.~2b we compare
the value of $E_{\rm sph}^{\rm MSSM}(T)$ (solid line) with the
corresponding scaling value given by Eq.~(\ref{scaling}). We can
see that the differences are $\simlt$ 5~\%~\cite{MOQ} which makes the use of
condition (\ref{condition}) reasonable.

\section{Baryogenesis in the MSSM}

Baryogenesis is fueled by CP-violating sources which are 
locally induced by the passage of the bubble wall~\cite{thick,thicknoi}.
These sources should
be inserted into a set of classical Boltzmann
equations describing  particle
distribution densities and permitting to take into account
Debye screening of induced
gauge charges~\cite{deb},
particle number changing reactions~\cite{cha} and
to trace the crucial role
played by diffusion~\cite{tra}. Indeed,
transport effects  allow  CP-violating charges to  efficiently
diffuse in
front of the advancing bubble wall where anomalous electroweak
baryon
violating processes are unsuppressed.

Following~\cite{newmethod1,newmethod2},
we are interested in the generation of charges which are
approximately conserved in the symmetric phase, so that they
can efficiently diffuse in front of the bubble where baryon number
violation is fast, and non-orthogonal to baryon number,
so that the generation of a non-zero baryon charge is energetically
favoured. Charges with these characteristics
are the axial stop ($\widetilde{t}$) charge 
and the Higgsino ($\widetilde{H}$) charge,
which may be produced from the interactions of squarks and
charginos and/or neutralinos with the bubble wall,
provided a source of CP-violation is present in these sectors.
CP-violating sources $\gamma_Q(z)$
(per unit volume and unit time) of a generic charge density
$J^0$ associated with  the current $J^\mu(z)$
and accumulated by the moving wall at a point
$z^\mu$ of the plasma can then be constructed from
$J^\mu(z)$~\cite{Toni} as $\gamma_Q(z)=\partial_0 J^0(z)$.

The detailed calculation of $\gamma_{\widetilde{q}}$ and 
$\gamma_{\widetilde{H}}$ has been recently performed~\cite{bau}.
It was proven that $\gamma_{\widetilde{q}}\ll
\gamma_{\widetilde{H}}$, due essentially to the chosen region in the
supersymmetric parameter space. Moreover, we have found that the Higgsino
current is given by
\begin{equation}
\label{current}
\langle J_{\widetilde{H}}^0(z)\rangle = \left| \mu\right| \sin\phi_{\mu}
\: \left[H^2(z) \Delta\beta/L_{\omega} \right]
\left[ 3 M_2 \; g_2^2 \; {\cal G}^{\widetilde{W}}_{\widetilde{H}}
 +       M_1 \; g_1^2 \; {\cal G}^{\widetilde{B}}_{\widetilde{H}}
\right],
\end{equation}
where
${\cal G}^{\widetilde{W}(\widetilde{B})}_{\widetilde{H}}$ are
integrals over the momentum space of the corresponding Feynman
diagrams, $\Delta\beta$ is the variation of the angle $\beta$
through the bubble wall and $L_{\omega}$ is the bubble wall
thickness. The integrand of 
${\cal G}^{\widetilde{W}(\widetilde{B})}_{\widetilde{H}}$
depends on the masses $\mu$, $M_2$ and $M_1$, as well as on the
temperature and on the widths (damping rates) that are taken
to be 
$\Gamma_{\widetilde{H}}\sim\Gamma_{\widetilde{W}}\sim\Gamma_{\widetilde{B}}\sim\alpha_W T$.

We can now solve the set of coupled differential
equations describing the effects of diffusion, particle number
changing reactions and CP-violating source terms.
We will closely follow the approach taken in
Ref.~\cite{bau} where the
reader is referred to for more details. The final
baryon-to-entropy ratio is found to be given by,
\begin{equation}
\frac{n_B}{s}=-g(k_i)\frac{{\cal A}\overline{D}\Gamma_{{\rm ws}}}
{v_{\omega}^2 s},
\label{baryon}
\end{equation}
where $v_{\omega}$ is the  wall velocity,
\begin{equation}
\label{higgs2}
{\cal A}=
\frac{1}{\overline{D} \; \lambda_{+}} \int_0^{\infty} du\;
\widetilde \gamma(u)
e^{-\lambda_+ u},
\end{equation}
$\overline{D}$ is the effective diffusion constant, 
\begin{equation}
\lambda_{+} = \frac{ v_{\omega} +
\sqrt{v_{\omega}^2 + 4 \widetilde{\Gamma}
\overline{D}}}{2 \overline{D}},
\end{equation}
$\widetilde{\Gamma}$ is the effective  decay constant, 
$\widetilde \gamma(z) = v_{\omega} \partial_{z}
J^0(z) f(k_i)$, and $f(k_i),g(k_i)$ are numerical
coefficients depending upon the light degrees of freedom.

\begin{figure}[htb]
\centerline{
\psfig{figure=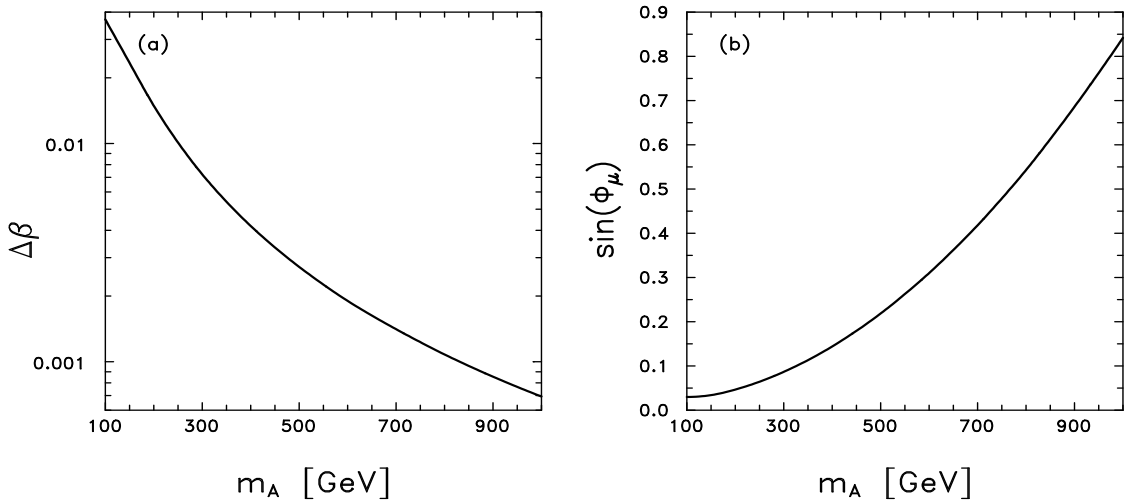,height=7.5cm,width=6cm,bbllx=9.cm,bblly=3.5cm,bburx=18.5cm,bbury=16.5cm}}
\caption{For $v_{\omega}=0.1$, $L_{\omega}=25/T$, $M_2=M_1=100$
GeV, $m_Q=500$ GeV, $\tan\beta=2$ and
$\widetilde{m}_U=\widetilde{m}_U^{\rm  crit}$: {\bf a)} Plot of
$\Delta\beta$ as a function of $m_A$. {\bf b)} Plot of $\sin\phi_{\mu}$
by fixing $n_B/s=4\times 10^{-11}$ (its lower bound).}
\end{figure}

From Eq.~(\ref{baryon}) one can see that the whole effect is
proportional to $\Gamma_{\rm ws}\sim 6\kappa \alpha_w^4\; T$, the weak
sphaleron rate in the symmetric phase. We have taken
$\kappa\sim 1$~\cite{AK} although its precise value is at present under
debate~\cite{ASY}.
We can also see from Eq.~(\ref{current}) that the final
baryon-to-entropy ratio depends on the parameter $\Delta\beta$.
This parameter should go to zero as $m_A\rightarrow\infty$ and
triggers the necessity of considering not too large values of $m_A$.  
We present in Fig.~3a a plot of $\Delta\beta$ as a function of $m_A$ which
confirms our expectatives. In Fig.~3b we plot $\sin\phi_{\mu}$
versus $m_A$ by fixing the value of $n_B/s$ to its lower bound 
$4\times 10^{-11}$ for the case $M_2=M_1=100$ GeV. The  values of the
effective diffusion and decay constants are $\overline{D}\sim 0.8\ {\rm
GeV}^{-1}$, $\widetilde{\Gamma}\sim 1.7$ GeV. We see, as
anticipated, that for large values of $m_A$, $\Delta\beta$
becomes very small and, correspondingly, $\sin\phi_{\mu}$
approaches 1. 

We conclude, from Fig.~3b, that the phase $\phi_{\mu}$ is never much
smaller than 0.05. These relatively large values of the phases are only
consistent with the constraints from the electric dipole moment of the
neutron if the  squarks of the first and second generation have masses
of the order of a few TeV~\cite{CKNlast}. Moreover, the baryon asymmetry
is not washed out inside the bubbles provided that the light stop is
lighter than the top quark, the pseudoscalar Higgs boson heavier than
$\sim$ 130 GeV and the lightest Higgs boson lighter than $\sim$ 80 GeV.

\section{Acknowledgements}
Work supported in part by the European Union (contract CHRX-CT92-0004) 
and CICYT of Spain (contract AEN95-0195). I wish to thank my 
collaborators A.~Brignole, M.~Carena, J.R.~Espinosa, A.~Riotto, I. Vilja,
C.~Wagner and F.~Zwirner.


\section{References}
\begin{thebibliography}{9}
%
\bibitem{baryogenesis} A.D.~Sakharov, {\it JETP Lett.} {\bf 91B}
(1967) 24.
%
\bibitem{reviews} For recent reviews, see:
A.G. Cohen, D.B. Kaplan and A.E. Nelson, 10 
{\it Annu. Rev. Nucl. Part. Sci.} {\bf 43} (1993) 27;
M. Quir{\'o}s, \HPA{67}{94}{451}; V.A.~Rubakov and
M.E.~Shaposhnikov, e-print [hep-ph/9603208]. 
%
\bibitem{anomaly}G.~t'Hooft, \PRL{37}{76}{8}; \PRD{14}{76}{3432}.
%
\bibitem{first} M. Shaposhnikov,
{\it JETP Lett.} {\bf 44} (1986) 465; \NPB{287}{87}{757} and
{\bf B299} (1988) 797.
%
\bibitem{improvement} M.E. Carrington, \PRD{45}{92}{2933};
M. Dine, R.G. Leigh, P. Huet, A. Linde and D. Linde,
\PLB{283}{92}{319};
\PRD{46}{92}{550}; P. Arnold, \PRD{46}{92}{2628};
J.R. Espinosa, M. Quir{\'o}s and F. Zwirner, \PLB{314}{93}{206};
W. Buchm{\"u}ller, Z. Fodor, T. Helbig and D. Walliser,
\AP{234}{94}{260}.
%
\bibitem{twoloop} J.~Bagnasco and M.~Dine, \PLB{303}{93}{308};
P. Arnold and O. Espinosa, \PRD{47}{93}{3546}; Z. Fodor and A.
Hebecker,
\NPB{432}{94}{127}.
%
\bibitem{nonpert} K. Kajantie, K.~Rummukainen
and M.E.~Shaposhnikov, \NPB{407}{93}{356};
 Z. Fodor, J. Hein, K. Jansen, A. Jaster and
I. Montvay, \NPB{439}{95}{147};
K.~Kajantie, M.~Laine, K.~Rummukainen and
M.E.~Shaposhnikov, \NPB{466}{96}{189};
K.~Jansen, e-print [hep-lat/9509018].
%
\bibitem{fs} G.R.~Farrar and M.E.~Shaposhnikov,
\PRL{70}{93}{2833},
({\bf E}): {\bf 71} (1993) 210 and \PRD{50}{94}{774}.
%
\bibitem{huet}
M.B. Gavela et al., \MPLA{9}{94}{795};
\NPB{430}{94}{382}; P.~Huet and E.~Sather, \PRD{51}{95}{379}.
%
\bibitem{early} G.F. Giudice, \PRD{45}{92}{3177};
S. Myint, \PLB{287}{92}{325}.
%
\bibitem{mariano1} J.R. Espinosa, M. Quir{\'o}s and F. Zwirner,
\PLB{307}{93}{106}.
%
\bibitem{mariano2} A. Brignole, J.R. Espinosa, M. Quir{\'o}s and F.
Zwirner,
\PLB{324}{94}{181}.
%
\bibitem{CQW} M. Carena, M. Quiros and C.E.M. Wagner,
\PLB{380}{96}{81}.
%
\bibitem{Delepine} D. Delepine, J.M. Gerard, R. Gonzalez Felipe
and J. Weyers, \PLB{386}{96}{183}.
%
\bibitem{MOQ} J.M.~Moreno, D.H.~Oaknin and M.~Quir{\'o}s,
\NPB{483}{97}{267},
and [hep-ph/9612212] to appear in {\em Phys. Lett.} {\bf B}.
%
\bibitem{CEQW} M. Carena, J.R. Espinosa, M. Quir{\'o}s and C.E.M.
Wagner,
\PLB{355}{95}{209}; M. Carena, M. Quir{\'o}s and C.E.M. Wagner,
\NPB{461}{96}{407}; H.E. Haber, R. Hempfling and A.H. Hoang, e-print
[hep-ph/9609331]. 
%
\bibitem{JoseR} J.R. Espinosa, \NPB{475}{96}{273}; B. de Carlos
and J.R.~Espinosa, e-print [hep-ph/9703212].
%

\bibitem{CE} D.~Comelli and J.R.~Espinosa, e-print [hep-ph/9606438].
%
\bibitem{BKS} A.I.~Bochkarev, S.V.~Kuzmin and M.E.~Shaposhnikov,
\PRD{43}{91}{369}. 
%
\bibitem{LEPRep} M. Carena, P. Zerwas and the Higgs Physics
Working
Group, in Vol. 1 of Physics at LEP2, G. Altarelli, T. Sj{\"o}strand
and F. Zwirner, eds., Report CERN 96-01, Geneva (1996).
%
\bibitem{thick} L. Mc Lerran {\it et al.}, {\it Phys. Lett.} {\bf B256}
(1991)
451; M. Dine, P. Huet and R. Singleton Jr.,
{\it Nucl. Phys.} {\bf B375} (1992)
625; M. Dine {\it et al.}, {\it Phys. Lett.} {\bf B257} (1991) 351;
A.G. Cohen and A.E. Nelson, {\it Phys. Lett.} {\bf B297} (1992) 111.
%
\bibitem{thicknoi}
D. Comelli, M. Pietroni and A. Riotto,
{\it Phys. Lett.} {\bf B354} (1995) 91
and {\it Phys. Rev.} {\bf D53} (1996) 4668.
%
\bibitem{deb} S.Yu. Khlebnikov, {\it Phys. Lett.} {\bf B300} (1993)
376;
 A.G. Cohen, D.B. Kaplan, and A.E. Nelson,
{\it Phys. Lett.} {\bf B294} (1992)
57; J.M. Cline and K. Kainulainen,
{\it Phys. Lett.} {\bf B356} (1995) 19.
%
\bibitem{cha}  A.G. Cohen, D.B. Kaplan, and A.E. Nelson,
{\it Phys. Lett.} {\bf 336} (1994) 41.
%
\bibitem{tra} M. Joyce, T. Prokopec and N. Turok,
{\it Phys. Rev. Lett.} {\bf 75} (1995) 1695, (E): {\it ibidem} 3375;
D. Comelli, M. Pietroni and A. Riotto,
{\it Astropart. Phys.} {\bf 4} (1995) 71.
%
\bibitem{newmethod1} P. Huet and A.E. Nelson,
\PLB{355}{95}{229}.
%
\bibitem{newmethod2} P. Huet and A.E. Nelson,
\PRD{53}{96}{4578}.
%
\bibitem{Toni} A. Riotto, {\it Phys. Rev.} {\bf D53} (1996) 5834.
%
\bibitem{bau} M. Carena, M. Quir{\'o}s, A. Riotto, I. Vilja and
C.E.M. Wagner, e-print [hep-ph/9702409].
%
\bibitem{AK} J. Ambj{\o}rn and A. Krasnitz, \PLB{362}{95}{97}.
%
\bibitem{ASY} P. Arnold, D. Son and L.G. Yaffe, e-print [hep-ph/9609481].
%
\bibitem{CKNlast} A.G.~Cohen, D.B.~Kaplan and A.E.~Nelson,
\PLB{388}{96}{588}. 
\end{thebibliography}
\end{document}